\documentclass[twocolumn,showpacs,preprintnumbers,amsmath,amssymb]{revtex4}
\usepackage{epsfig}
\usepackage{dcolumn}
\usepackage{bm}

\begin{document}


\title{Antidot tunneling between Quantum Hall liquids with different
filling factors}

\author{V.V. Ponomarenko\footnote{on leave from St. Petersburg State
Polytechnical University, Center for Advanced Studies,
St. Petersburg 195251, Russia.} and D.V. Averin}

\address{Department of Physics and Astronomy, University of Stony
Brook, SUNY, Stony Brook, NY 11794}

\date{\today}


\begin{abstract}
We consider tunneling through two point contacts between
two edges of Quantum Hall liquids of different filling
factors $\nu_{0,1}=1/ (2m_{0,1}+1)$ with $m_0-m_1\equiv
m>0$. Properties of the antidot formed between the point
contacts in the strong-tunneling limit are shown to be
very different from the $\nu_0 =\nu_1$ case, and include
vanishing average total current in the two contacts and
quasiparticles of charge $e/m$. For $m>1$, quasiparticle
tunneling leads to non-trivial $m$-state dynamics of
effective flux through the antidot which restores the
regular ``electron'' periodicity of the current in flux
despite the fractional charge and statistics of
quasiparticles.
\end{abstract}


\maketitle

One of the most interesting features of the Fractional
Quantum Hall Liquids (FQHLs) is the existence of
quasiparticles with fractional charge \cite{b1} and
exchange statistics \cite{b2}. Although the
quasiparticles are defined most simply in the
incompressible bulk of the FQHL, in the low-energy
transport experiments, quasiparticles are created
typically at the liquid edges, e.g. by tunneling between
them. In the simplest case of FQHL with the filling
factor $\nu= 1/odd$, the quasiparticles that tunnel
through the liquid between its edges coincide with the
quasiparticles in the bulk \cite{b3} and their
fractional charge $\nu e$ can be measured experimentally
\cite{b4,b5}. So far, fractional statistics of
quasiparticle has not been directly observed in
experiments, although there is experimental \cite{b5*}
and theoretical \cite{b6} interest to manifestations of
this statistics in the noise correlators of the tunnel
currents.

Strong tunneling between edges of FQHLs with different
filling factors should create quasiparticles which are
different from those in the bulk of the liquids but
still have fractional charge and statistics
\cite{b7,b8}. Untill now, such tunneling has been
considered only in the geometry of a single point
contact \cite{b7} or multiple contacts \cite{b8} for
which the interference between different contacts is not
important (i.e., the edges do not form closed loops).
The purpose of this work is to study an ``antidot''
tunnel junction: two separate point contacts at points
$x_1$, $x_2$ along the $x$-axis between two single-mode
edges of QHLs with different filling factors
$\nu_{0,1}=1/(2 m_{0,1}+1)$ with $m_0>m_1\geq 0$ -- see
Fig.\ 1. In this geometry, the tunneling processes at
two point contacts interfere and statistics of tunneling
quasiparticles directly affects the dc current.

If the two filling factors are equal, $\nu_0= \nu_1
\equiv \nu$, as in experiments \cite{b4}, strong
tunneling leads to formation of a closed edge between
the points $x_1$ and $x_2$ encircling the antidot and
separated from external edges of the surrounding uniform
QHL \cite{b9}. Quasiparticles of charge $\nu e$ can then
tunnel between the external edges through the antidot.
As shown below, the situation is very different for
$\nu_0\neq \nu_1$, and the antidot formed between the
point contacts does not decouple completely from
external edges even in the limit of strong tunneling. As
a result, the total tunnel current between the two QHLs
vanishes in this limit. Also, interference between the
two contacts produces the quasiparticles of charge $e/m$
determined by the change $m$ of the number of flux pairs
attached to the electrons \cite{b10} in the two liquids:
$m=m_0-m_1$. In general, proper account of the flux
attachement to tunneling electrons is one of the key
elements of the description of strong multi-point
tunneling between QHLs with different filling factors.
Changes in the number of attached flux pairs should be
accounted for by inclusion of appropriate statistical
contributions into the bosonic tunneling fields
describing different point contacts \cite{b11}. These
statistical terms do not affect the perturbative
expansions in tunneling, but make the tunneling fields
at different point contacts commute, the feature that
becomes important in the limit of strong tunneling.

\begin{figure}[htb]
\setlength{\unitlength}{1.0in}
\begin{picture}(3.,1.26)
\put(0.85,0.02){\epsfxsize=1.6in \epsfbox{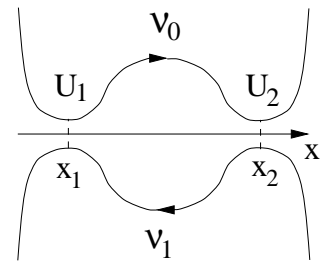}}
\end{picture}
\caption{``Antidot'' junction considered in this work:
two point contacts with tunneling amplitudes $U_j$
formed at points $x_j$, $j=1,2$, between two
counter-propagating edges of QHLs with different filling
factors $\nu_0$ and $\nu_1$. The edges are assumed to
support one bosonic mode each, with arrows indicating
direction of propagation of these modes.}
\end{figure}

Quantitatively, we describe the antidot junction
(Fig.~1) between two single-mode edges with filling
factors $\nu_l=1/ (2m_l+1)$, $l=0,1$, by adopting the
standard bosonization approach in which the electron
operator $\psi_l$ of the edge $l$ is expressed as
\cite{b12}
\[ \psi_l=(D/2\pi v_l)^{1/2} \xi_l e^{i(-1)^l [\phi_l(x,t)/
\sqrt{\nu_l} +k_lx] } . \] Here $\phi_l$ are the two
bosonic modes propagating in the opposite direction with
velocities $(-1)^l v_l$, $v_l>0$, the Majorana fermions
$\xi_l$ account for mutual statistics of electrons in
different edges, and $D$ is a common energy cut-off of
the edge modes. The Fermi momenta $k_l$ correspond to
the average electron density in the edges, while the
operators of the density fluctuations are:
$\rho_l(x,\tau)= (\sqrt{\nu_l}/2 \pi) \partial_x
\phi_l(x, \tau)$. Bosonic fields $\phi_l$ have the
standard quadratic Lagrangian defined by the Fourier
transform of the imaginary-time-ordered correlators
(see, e.g., \cite{b8}) $\langle \phi_l (x) \phi_p
\rangle= \delta_{lp} g((-1)^l x/v_l,\omega)$, where
\begin{equation}
g(z,\omega)=\frac{2 \pi}{\omega} \mbox{sgn} (z)
\Big(-{1\over 2}+\theta(\omega z)e^{-\omega z} \Big) \,
, \label{e1} \end{equation} which in particular imply
the usual equal-time commutation relations
$[\phi_l(x),\phi_p(0)]=i \pi \mbox{sgn}(x)\, \delta_{lp}
(-1)^l $.

With the bosonized electron operators, Langrangian
describing electron tunneling in the two contacts is:
\begin{equation}
{\cal L}_t = \sum_{j=1,2} [\frac{DU_j}{2\pi}
e^{i\kappa_j} e^{i \lambda \varphi_j} + h.c.] \equiv
\sum_{j=1,2} (T_j^+ +T_j^-)\, , \label{e2}
\end{equation} where $U_j$ and $\kappa_j$ are the
absolute values and the phases of the dimensionless
tunneling amplitudes, and
\[ \lambda \varphi_j(t) \equiv {\phi_0(x_j,t) \over
\sqrt{\nu_0}} + {\phi_1(x_j,t) \over \sqrt{\nu_1} } \, ,
\;\;\; \lambda=\left[ {\nu_0 +\nu_1\over \nu_0 \nu_1
}\right]^{1/2} \, . \] The factor $\lambda$ is chosen in
such a way that the normalization of the bosonic
operators $\varphi_j$ coincides with that of the fields
$\phi_l$, so that the imaginary-time correlators of
$\varphi_j$ follow directly from Eq.~(\ref{e1}). The
products of the Majorana fermions $\xi_1 \xi_2$ were
omitted from the Lagrangian (\ref{e2}), since they at
most can produce an irrelevant overall constant shift of
the phases $\kappa_j$. These phases include
contributions from the external magnetic flux $\Phi$
through the antidot and from the average electron
numbers $N_{0,1}$ on the two sides of its perimeter, so
that $\kappa_2 - \kappa_1=2\pi[(\Phi/\Phi_0)+
(N_0/\nu_0)+(N_1/\nu_1)]+ \mbox{const} \equiv \kappa $,
where $\Phi_0=h/e$ is the ``electron'' flux quantum.

If the bias voltage $V$ is applied to the junction, the
electron current operator is
$I^{e}=i\sum_{j=1,2}\sum_{\pm} (\pm) T_j^{\pm} e^{\mp
iVt}$. Its average includes contributions from the
individual point contacts $\bar{I}^e_j$ and the
phase-sensitive interference term $\Delta I^e(\kappa)$:
$\langle I^e \rangle =\sum_j \bar{I}^e_j+\Delta
I^e(\kappa)$. At temperature $T$, in the lowest
non-vanishing order of the perturbation theory in $U_j$,
the two contributions are:
\begin{equation}
\bar{I}^e_j= (U^2_jD/2\pi) (2\pi T/D)^{\lambda^2-1}
C_{\lambda^2}(V/2\pi T)\, , \label{e3} \end{equation}
where $C_g (v) \equiv \sinh (\pi v)|\Gamma (g/2 +iv)|^2/
[\pi\Gamma(g)]$, and
\[ \Delta I^e= ({2U_1U_2D\over \pi}) ({2\pi T \over D})^{\lambda^2
-1} \mbox{Im} \Big\{ \int^\infty_{-\infty} ds \sin
(\kappa- {sV\over \pi T}) \]

\vspace{-3ex}

\begin{equation}
\cdot \prod_{l=0,1} [i\sinh(s-(-1)^lt_l\pi
T-i0)]^{-1/\nu_l} \Big\} \, . \label{e4} \end{equation}
Here $t_l$ is the time of electron propagation between
the two point contacts along the $l$th edge.

Equations (\ref{e3}) and (\ref{e4}) show that at low
energy, when $V,T< \Delta$, where $\Delta
=t_{\Sigma}^{-1}$, $t_{\Sigma}\equiv t_0+t_1$, the terms
$t_l\pi T$ can be omitted in Eq.\ (\ref{e4}), and the
full current $\langle I^e \rangle$ is given by Eq.\
(\ref{e3}) with the tunnel amplitude $U_j$ replaced by
the coherent sum of the two point-contact amplitudes:
$U_j\rightarrow |U_1e^{i\kappa_1}+U_2e^{i\kappa_2 }|$.
This implies that similarly to the situation of one
point contact, the total electron tunneling current
$\langle I^e \rangle$ is suppressed at low energies $E$
roughly as $E^{\lambda^2-1}$, where $E\simeq \mbox{max}
(V,T)$. At large energies, the total current $\langle
I^e \rangle$ is dominated by the contributions
(\ref{e3}) from individual point contacts, with
interference term (\ref{e4}) suppressed exponentially at
large temperatures $T>\Delta$, or growing slower that
the individual currents (\ref{e3}) with increasing
voltage $V>\Delta>T$. Therefore, if $\Delta< T_X$,
transition from weak to strong tunneling occurs at the
same energy scale $T_X \simeq D U^{-2/ (\lambda^2-2)}$
(we assume for simplicity that $U_1 \simeq U_2 \equiv
U$) as for one point contact \cite{b13,b14}, with the
perturbative result for electron tunneling valid at
$T,V<T_X$. As shown below, if $\Delta> T_X$, electron
tunneling become strong and perturbation theory in $U$
is not correct.

The main focus of our work is on this {\em
strong-tunneling limit} that can be realized if $U\gg
1$. In this case, the tunneling Lagrangian (\ref{e2})
gives the dominant part of the action and a natural
approach to solving the systems dynamics would be to fix
the tunneling modes $\varphi_j$ at the extrema of the
Lagrangian (\ref{e2}). This, however, can not be done
directly since the operators $\varphi_j$ at two point
contacts do not commute: $[\varphi_1,\varphi_2]=i \pi
(\nu_1-\nu_0)/ (\nu_1+ \nu_0)$, and can not be
simultaneously fixed, although the different transfer
terms $T_j^{\pm}$ (\ref{e2}) still commute among
themselves, e.g.,
\begin{equation}
T_1^{\pm}T_2^{\pm}=e^{2\pi m i} T_2^{\pm} T_1^{\pm} \, .
\label{e5} \end{equation}

This problem can be resolved using the fact that the
tunneling Lagrangian (\ref{e2}) does not uniquely define
the strong-tunneling limit. The factors $\exp\{\pm
i\sqrt{2m} \eta_j\}$ with free zero-energy bosonic modes
$\eta_j$ defined by their imaginary-time-ordered
correlators: $<T_\tau{\eta_i(\tau) \eta_j(0)}>=i \pi
\Theta((j-i)\tau) (1-\delta_{ij})$, can be included
\cite{b11} into the terms $T_j^{\pm}$ in (\ref{e2})
without changing the perturbation expansion of the
partition function in ${\cal L}_t$ in any order. The new
tunneling operators $\varphi_j+ \sqrt{2m} \eta_j$
commute, and can be fixed by the natural
strong-tunneling conditions:
\begin{equation}
\lambda \varphi_j+ \sqrt{2 m} \eta_j+\kappa_j=2 \pi n_j.
\label{e5*} \end{equation}

Transport properties of the antidot junction (Fig. 1)
can be described then as successive transformations of
the fields $\phi^{in}(x_j)$ entering individual contacts
into the outgoing fields $\phi^{out}(x_j)$:
$\phi^{out}(x_j)=\hat{P} \phi^{in}(x_j)$, where
\cite{b7,b8}
\begin{equation}
P_{00}=-P_{11} = {\nu_0- \nu_1\over \nu_0+ \nu_1} \, ;
\;\; P_{01}=P_{10} = -{2\sqrt{\nu_0 \nu_1} \over \nu_0
+\nu_1} \, , \label{e7} \end{equation} and $\phi^{in,\,
out} (x_j)= (\phi_0 (x_j\mp 0), \phi_1(x_j\pm 0))^T$.
The transformations (\ref{e7}) at the two contacts
should be combined according to the edge propagation
diagram shown in Fig.\ 2, which involves multiple
interferences along the closed loop formed between the
points $x_1$ and $x_2$. Summation over these
interferences gives a very simple final result
$\phi^{out}=\phi^{in}$ which means that the fields and
the currents incident on the junction remain the same in
the outgoing edges, and the total tunnel current between
the two QHL vanishes.

\begin{figure}[htb]
\setlength{\unitlength}{1.0in}
\begin{picture}(3.3,.7)
\put(0.04,0.04){\epsfxsize=3.1in \epsfbox{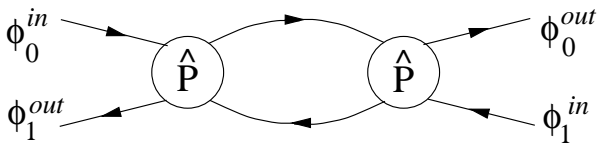}}
\end{picture}
\caption{Diagram of the strong-coupling edge propagation
in the antidot junction.}
\end{figure}

This discussion assumes that large tunnel amplitudes
$U_j$ completely suppress quantum fluctuations of the
fields $\varphi_j$. For large but finite $U_j$, the
fields can tunnel between different minima $n_j$ of the
strong-coupling solution (\ref{e5*}), the process that
represents quasiparticle backscattering and can be
described quantitatively in terms of the instanton
expansion \cite{b8}. The operators $(W_jD/2\pi) \exp \{
\pm i 2 \theta_j/ \lambda \}$, where $\theta_j$ are dual
to $\varphi_j$: $[\theta_j, \varphi_k]= i\pi (-1)^j
\delta_{jk}$, generate the jumps of $\varphi_j$ changing
$n_j$ by $\mp(-1)^j$. Instanton tunneling amplitude
$W_j$ can be estimated as in one contact, $W_j\simeq
U_j^{[(2/\lambda^2)-1]/ [(\lambda^2/2)-1]}$ \cite{b16}.
Using the commutation relations of the edge fields
$\phi_l$ one can express $\theta_j$ in terms of
$\phi_l$:
\begin{equation}
\theta_{j} = (-1)^{j+1} [\phi_0(x_j-0)/\sqrt{\nu_0} +
\phi_1(x_j+ 0)/\sqrt{\nu_1}]/ \lambda  \, , \label{e8}
\end{equation} and see that as in the case of one point
contact \cite{b14}, instanton tunneling operators
transfer charge $\pm (-1)^j 2\nu_0 \nu_1/ (\nu_0+
\nu_1)$ from QHL$_{\nu_1}$ into the QHL$_{\nu_0}$. In
the antidot junction, however, these charges undergo
further scattering in accordance with the diagram in
Fig.\ 2. Accounting for this scattering with the help of
Eq.~(\ref{e7}) we see that the total quasiparticle
charge transferred from QHL$_{\nu_1}$ into the
QHL$_{\nu_0}$ by each instanton tunneling is $\pm 1/m$.

Evaluating action on the instanton trajectories
similarly to what is done in \cite{b8}, we obtain the
Lagrangian $\bar{{\cal L}}_t$ for quasiparticle
tunneling that has the form dual to the Lagrangian
(\ref{e2}):
\[ \bar{{\cal L}}_t=\sum_{j=1,2}\Big[ {W_jD\over 2\pi } \bar{F}_j
\exp \big\{ i \big({\kappa_j \over m} +{2\theta_j \over
\lambda} - {Vt \over m}\big)\big\}  + h.c.\Big]  \]

\vspace{-3 ex}

\begin{equation}
\equiv \sum_{j=1,2} \sum_{\pm} \bar{T}_j^{\pm} e^{\mp
iVt /m} , \label{e10} \end{equation} The correlators of
the fields $\theta_{j}$ in (\ref{e10}) can be found from
Eq.~(\ref{e8}) combined with the strong-coupling
scattering scheme described by Fig.\ 2 and
Eq.~(\ref{e7}):
\begin{equation}
\langle \theta_j \theta_j \rangle = g(0,\omega)+ {1
\over 2} \sum_{n=1}^\infty P_{01}^{2n} \sum_\pm g(\pm
nt_\Sigma, \omega), \label{e9} \end{equation}

\vspace{-3ex}

\[ \langle \theta_1\theta_2 \rangle= \sum_{n=0}^\infty
\frac{ P_{01}^{2n} }{\nu_0+\nu_1} [\nu_1 g(t_0 + n
t_\Sigma, \omega) +\nu_0 g(-t_1-n t_\Sigma, \omega)] .\]

The unitary operators $\bar{F}_j$ in (\ref{e10}) are
Klein factors describing statistics of quasiparticles
and are characterized by the following relations:
\begin{equation}
\bar{F}_1\bar{F}_2=e^{2 \pi i\over m} \bar{F}_2\bar{F}_1
, \;\;\; \langle \bar{F}^k_1 (\bar{F}^+_1)^p\bar{F}^l_2
(\bar{F}^+_2)^q \rangle =\delta_{kp}\delta_{lq}\, ,
\label{e12} \end{equation} were the Kronecker symbol
$\delta_{ij}$ is defined modulo $m$. These relations
originate from the effective flux through the antidot
which includes external magnetic flux and statistical
contribution and is equal to $n=n_1-n_2$ (in units of
$\Phi_0$) in the strong-tunneling limit (\ref{e5*}).
Because of the interference in the two point contacts,
tunneling amplitudes of quasiparticles of charge $1/m$
acquire the phases $e^{\pm i 2\pi n/m}$ which
distinguish $m$ states of the antidot with different
fluxes $n'\neq n\, \mbox{mod} (m)$. Since these states
are equivalent in all other respects, summation of the
tunneling amplitudes over them gives rise to the second
relation in (\ref{e12}). On the other hand, each
quasiparticle tunneling changes $n$ by $\pm 1$, the fact
accounted for by the first part of Eq.~(\ref{e12}),
which also ensures commutativity of the transfer
operators $\bar{T}_j^{\pm}$ in (\ref{e10}). In addition,
electron commutation relations (\ref{e5}) mean that each
electron tunneling changes the effective flux between
the two tunneling points $x_{1,2}$ by $m$, and the
variation of the flux by $\pm 1$ due to quasiparticle
tunneling can be seen as the basic reason for the
fractional quasiparticle charge $1/m$.

From Lagrangian (\ref{e10}), the operator $I$ of the
quasiparticle current is: $I=(i/m)
\sum_{j=1,2}\sum_{\pm} \pm \bar{T}_j^{\pm} e^{\mp
iVt/m}$. Similarly to electron current (\ref{e3}),
(\ref{e4}), its average contains contributions from
individual point contacts and phase-sensitive
interference term: $\langle I \rangle = \sum_j
\bar{I}_j+\Delta I (\kappa)$. Since in the absence of
backscattering the current between the two QHLs vanishes
in the strong-tunneling limit, quasiparticle
backscattering gives the full current. Behavior of this
current depends on the relation between the crossover
energy $T_X$ and the antidot ``quantization'' energy
$\Delta$. If the two point contacts are far apart and/or
the tunneling amplitudes $U_j$ are sufficiently small,
then $\Delta <T_X$, and the transition from weak to
strong tunneling occurs at large energies at which the
interference is already suppressed, so that this
transition has the same form as in one point contact
\cite{b13,b14}. In particular, at energies above $T_X$,
the lowest-order perturbation theory in quasiparticle
tunneling (\ref{e10}) gives:
\begin{equation}
\bar{I}_j={1\over m}(W^2_jD/2\pi) \big({2\pi T \over D}
\big)^{4/\lambda^2-1} C_{4/\lambda^2}(V/2\pi T m) \, .
\label{e13} \end{equation} The current (\ref{e13}) is
carried by individual quasiparticles of charge $1/m$,
which should be seen at $V\gg T$ in the shot noise
caused by the flow of this current.

If the two point contacts are close and/or $U_j$'s are
large, $\Delta >T_X$, and electron interference in the
two contacts becomes important. Renormalized electron
tunneling amplitude for the current encircling the
antidot can be estimated roughly as $U
(\Delta/D)^{\lambda^2 -1}$ and becomes large when
$\Delta >T_X$. In this regime, dynamics of tunneling
should be discussed in terms of quasiparticle
backscattering even at low energies $V,T$. The form of
transition from weak to strong interference at $\Delta
\simeq T_X$ depends on $m$ which controls the energy
scaling of the quasiparticle amplitudes $W_j$ at low
energies. As one can see from the correlators (\ref{e9})
of the fields $\theta_j$, the scaling dimensions of
$W_j$ at energies above and below $\Delta$ are equal to
$2/\lambda^2 -1$ and $\lambda^2/2m^2 -1$, respectively.
In the case of {\em one-flux-pair variation}, $m=1$,
(i.e., $\nu_0=1/3$ and $\nu_1=1$, or $\nu_0=1/5$ and
$\nu_1=1/3$) both the quasiparticle charge and the
low-energy scaling dimension $\lambda^2/2 -1$ of
tunneling amplitudes coincide with those for electrons.
This means that for $T_X<\Delta$ the quasiparticle
current is perturbative in $W$ at any $V,T$. Below
$\Delta$ it has the same $V,T$-dependence and similar
interference pattern, $\langle I \rangle \propto |W_1
e^{i\kappa_1} +W_2e^{i\kappa_2 }|^2$, as electron
current in Eqs.~(\ref{e3}) and (\ref{e4}) for
$V,T<\Delta$. The relation between $\Delta$ and $T_X$
affects then only the $U$-dependence of the overall
magnitude of the current, which increases or decreases
with increasing $U$ in the electron and quasiparticle
regimes, respectively. The two currents coincide at
$\Delta \simeq T_X$.

In the case of the {\em two-flux-pair variation}, $m=2$,
realized when $\nu_0=1/5$ and $\nu_1=1$, the Klein
factors (\ref{e12}) anticommute and can be represented
by Pauli matrices: $\bar{F}_ {1,2}=\sigma_{1,2}$. In
terms of the flux $n$ through the antidot, the Pauli
matrices $\sigma$ act in the space of two states
representing the parity of $n$. The low-energy scaling
dimension of the quasiparticle tunneling operators
$\bar{T}_{1,2}^{\pm}$ is equal to $-1/4$, the fact that
makes Lagrangian (\ref{e10}) non-perturbative at low
energies $V,T$. To understand this non-perturbative
behavior, we notice that for $V,T< \Delta$, the
quasiparticle terms $\bar{T}_{1,2}^{\pm}$ can be reduced
to effective single-point tunneling
\begin{equation}
\bar{T}_1^++\bar{T}_2^+= (\Delta /4 \pi) e^{i {\lambda
\over 2} \vartheta(t) } \sum_\pm  X_\pm e^{i
\varphi_\pm} \sigma_\pm \, , \label{e14} \end{equation}

\vspace{-4ex}

\[ X_\pm \simeq (\Delta/D)^{2/\lambda^2-1} [W_1^2+W_2^2 \mp
2W_1W_2\sin(\kappa/2)]^{1/2} \, ,\] where
$\varphi_\pm=\mbox{arg} [W_1 \mp iW_2 e^{-i\kappa/ 2}]$,
free bosonic field $\vartheta$ is defined by the
correlator $\langle \vartheta(t)\vartheta(0) \rangle
=g(0,t)$, and $\sigma_\pm= (\sigma_1\pm i\sigma_2)/2$.
The model (\ref{e14}) can be mapped onto a model
\cite{b13} of zero-energy resonant level in the
Tomonage-Luttinger liquids (TLL) of interaction constant
$g=1/3$. In this mapping, the amplitudes of tunneling
into the level from the two TLL electrodes are
$\sqrt{\Delta/2 \pi}X_\pm$, and our quasiparticle
current coincides with the current through the level.
Under conditions of resonance, $\kappa=2\pi \times
integer$, the two amplitudes $X_{\pm}$ are equal, and
the model (\ref{e14}) can be further reduced to a point
scatterer in the uniform TLL of interaction strength
$g^{-1}/4=3/4$. The quasiparticle current $\langle I
\rangle$ then is:
\begin{equation}
\langle I \rangle =(\sigma_0 V/3)[1-G_{3/4}(2V/3 \Gamma,
T /\Gamma)] , \label{e15} \end{equation} where $\Gamma
\propto \Delta X^4 \simeq \Delta (T_X/ \Delta)^{8/3}$,
and $\sigma_0$ is the free electron conductance. The
function $G_g$ is normalized TLL conductance and
describes the cross-over from $G_g(v,0) \simeq (1 -
c_1(1/g) v^{2(1/g-1)})$ for $v<1$ to $G_g(v,0) \simeq
c_1(g) v^{2(g-1)}$ for $v>1$ at zero temperature $T$,
where $c_1(g)=\sqrt{\pi} \Gamma(g+1)/ (2
\Gamma(1/2+g))$, and similar dependence on $T$ at zero
voltage $V$. The maximum conductance $\sigma_0 /3$
(\ref{e15}) is reached at $V,T \ll \Gamma$. The
short-noise charge is equal to $2/3$ and is twice larger
than the charge of one TLL quasiparticle, since only
backscattering of pairs of quasiparticles is allowed at
resonance. Small deviations of $\kappa$ from the
resonant values restore backscattering of individual
quasiparticles and lead to suppression of the tunnel
current outside the resonances. The width of resonances
changes with temperature as $(T/\Delta)^{2/3}$.

In conclusion, we have developed a theory of transport
properties of antidot junction between two single-mode
edges of different filling factors (Fig.~1). The theory
predicts tunneling of quasiparticles of fractional
statistics and charge $e/m$ set by variation $m$ of the
number of flux pairs attached to electrons in the QHLs.
The charge $e/m$ should be seen in the shot noise for
weak interference between backscattering processes in
the two point contacts of the junction, while
quasiparticle statistics affects the tunnel current when
the interference is strong.

This work was supported by the NSA and ARDA under ARO
contract \# DAAD19-03-1-0126.

\end{document}